\documentclass{emulateapj}
\usepackage{apjfonts}
\usepackage{graphicx}
\usepackage{amssymb}
\usepackage{dsfont}
\usepackage{color} 
\usepackage{textcomp}

\newcommand{\ron}{}  \newcommand{\strike}[1]{}
\newcommand{\bon}{} 

  \submitted{Accepted for publication in The Astrophysical Journal}

\begin{document} 

\bibliographystyle{apj}

\shortauthors{Carter, Rappaport,  \& Fabrycky}
\shorttitle{{\em Kepler} Hot White Dwarf Companion}
\title{A Third Hot White Dwarf Companion Detected by {\em Kepler}}

\author{Joshua A. Carter\altaffilmark{1,2,3}, Saul Rappaport\altaffilmark{1}, \& Daniel Fabrycky\altaffilmark{2,3,4}}
\email{jacarter@cfa.harvard.edu}

\ron{ \altaffiltext{1}{M.I.T. Kavli
 Institute for Astrophysics and Space Research, 70 Vassar St.,
 Cambridge, MA, 02139} 
 \altaffiltext{2}{Harvard-Smithsonian CfA, 60 Garden St.,
 Cambridge, MA, 02138} 
  \altaffiltext{3}{Hubble Fellow}}
 \altaffiltext{4}{UCO/Lick, University of California, Santa Cruz, CA 95064}

\begin{abstract}
	We have found a system listed in the {\em Kepler} Binary Catalog ($P_{\rm orb} = 3.273$ days; Prsa et al. 2010) that we have determined is comprised of a low-mass, thermally-bloated, hot white dwarf orbiting an A star of about 2.3 $M_\odot$.  In this work we designate the object, KIC 10657664, simply as \ron{``}KHWD3\ron{''} \ron{(``Kepler Hot White Dwarf 3'')}. We use the transit depth of $\sim$0.66\%, the eclipse depth of $\sim$1.9\%, and regular smooth periodic variations at the orbital frequency and twice the orbital frequency to analyze the system parameters.  The smooth periodic variations are identified with the classical ellipsoidal light variation (``ELV'') and illumination (``ILL'') effects, and the newly utilized Doppler boosting (``DB'') effect. Given the measured values of $R/a$ and inclination angle of the binary, both the ELV and DB effects are mostly sensitive to the mass ratio, $q = M_2/M_1$, of the binary. The two effects yield values of $q$ which are somewhat inconsistent -- presumably due to unidentified systematic effects -- but which nonetheless provide a quite useful set of possibilities for the mass of the white dwarf (either \strike{$0.18 \pm 0.03 \,M_\odot$ or $0.37 \pm 0.08 \, M_\odot$}\ron{$0.26 \pm 0.04 \,M_\odot$ or $0.37 \pm 0.08 \, M_\odot$}).  All of the other system parameters are determined fairly robustly.  In particular, we show that the white dwarf has a radius of $0.15 \pm 0.01 \, R_\odot$ which is extremely bloated over the radius it would have as a fully degenerate object, and an effective temperature \strike{$T_{\rm eff} = 14,100 \pm 350$}\ron{$T_{\rm eff} \simeq 14,500$} K.  Binary evolution scenarios and models for this system are discussed.  We suggest that the progenitor binary was comprised of a primary of mass $\sim$2.2 $M_\odot$ (the progenitor of the current hot white dwarf) and a secondary of mass $\sim$1.4 $M_\odot$ (the progenitor of the current A star in the system).  We compare this new system with three other white dwarfs in binaries that likely were formed via stable Roche-lobe overflow (KOI-74, KOI-81, and \strike{Regulus}\bon{the inner Regulus binary}).
\end{abstract}


\maketitle

\section{Introduction}

The exquisite photometric precision of the {\em Kepler} mission has led, among many other things, to the discovery of two systems in which a hot white dwarf both transits its parent star, and is eclipsed by it (Rowe et al.\,2010; van Kerkwijk et al.\,2010).  In the case of KOI-74, the relative transit and eclipse depths are only $5 \times 10^{-4}$ and $12 \times 10^{-4}$, respectively, while the amplitudes of the ellipsoidal light variations and Doppler boosting effect have even more remarkably small amplitudes of $1.4 \times 10^{-4}$ and $1.0 \times 10^{-4}$, respectively.    Such small effects are unexplored with ground-based astronomy.

The discovery of transiting/eclipsing white dwarfs which exhibit the effects mentioned above are amenable to many interesting system diagnostics, and can thereby be potentially very revealing about the binary stellar evolution that has led to their current configuration (e.g., van Kerkwijk et al.\,2010; Di Stefano 2010).  In particular, if the progenitor of the white dwarf was initially of mass $\lesssim 2.2\,M_\odot$ the orbital period may be tightly correlated with the white dwarf mass (e.g., Rappaport et al.\,1995).  In all cases where the current-epoch binary orbital period is short (e.g., less than a month) the system almost certainly involved a phase of mass transfer from the progenitor of the white dwarf to what is presently the normal stellar companion.  In the process, a substantial amount of the original primary mass was transferred to its companion or lost from the system.  The temperature of the white dwarf provides some important information on the cooling history and the age since the mass transfer event (see, e.g., van Kerkwijk et al.\,2010; Di Stefano 2010).

In this work, we report the discovery of a third hot white dwarf (with $T_{\rm eff} \simeq 14,500$ K) orbiting an A star, which we herein designate as ``KHWD3''.  The star, KIC 10657664, was identified and catalogued by Prsa et al. (2010) in the {\em Kepler} binary star catalog, although it was not identified as harboring a hot white dwarf companion.  Its transit and eclipse depths are about an order of magnitude larger than those of KOI-74 and KOI-81, due mostly to the fact that the radius of the white dwarf in KHWD3 is a factor of $\gtrsim$3 times larger than in the systems discovered earlier.  

The layout of this paper is as follows.  In \S2, we present the {\em Kepler} light curve for this system, and the analyses which yield the flux ratio of the two stars, the ratio of their radii, the orbital inclination angle, and the first three harmonics of the light curve which reveal ellipsoidal light variations, an illumination effect, and Doppler boosting.  In \S3 we determine the model-independent parameters of the binary system including the mass, effective temperature, radius, and luminosity of the white dwarf, and the mass ratio.  Inferences about the properties of the primary star allow us to estimate its mass at $2.3 \pm 0.4\, M_\odot$.  From this estimate and the measured mass ratio we arrive in \S 4 at two possible solutions for the mass of the white dwarf: either \strike{$\sim$$0.18\,M_\odot$ or $\sim$$0.37\,M_\odot$}\ron{$0.26 \pm 0.04 \,M_\odot$ or $0.37 \pm 0.08\,M_\odot$}, depending on whether we emphasize in our analysis the ellipsoidal light variations or the Doppler boosting effect.   In \S5 we discuss various evolutionary scenarios that can lead to the present system.

\nopagebreak{\section{Light Curve Analysis}  \label{sec:model}
The uncorrected light curve for KHWD3, downloaded from the {\em Kepler} public archive\footnote{{\url http://archive.stsci.edu/kepler/publiclightcurves.html}}, is shown in Figure~\ref{fig:lc}.  The data are comprised of $\sim10$ days of quarter zero (Q0) data and $\sim33$ days of quarter one (Q1) data\footnote{{\em Kepler} ``quarters'' are continuous observing blocks separated by short times in which data are downloaded from the spacecraft.  Only Q0 and Q1 are currently available to the public.  \ron{We have utilized only the unprocessed (``raw'') data products in our analysis.}}.  
Outside of the easily-identified eclipses, the variation of the light curve is dominated by a low frequency trend, assumed to be a systematic associated with the {\em Kepler} spacecraft.  The remaining periodic variation on times at or shorter than the orbital period and in phase with the conjunctions is assumed to be astrophysical in nature (as a result of, for example, ellipsoidal light variations and/or Doppler boosting).   We accounted for these three effects in a light curve model, as described below, and determined parameters of the eclipsing system and the harmonic content of the out-of-eclipse variation.}

\subsection{Light Curve Model}

We modeled the eclipse events assuming two spherical stars having radius ratio $R_2/R_1$, and observed flux ratio $F_2/F_1$, assigning the indices of 1 and 2 to primary and secondary, respectively.  The binary was constrained to a Keplerian orbit parameterized by a period $P$, a normalized semi-major axis distance $a/R_1$, an inclination to the sky plane $i$, an eccentricity $e$, an argument of periastron $\varpi$ and a time of inferior conjunction (i.e., at the time of the eclipse of the primary), $t_{\rm ic}$. 

The normalized eclipse light curve, $f(t)$, was calculated to be
\begin{eqnarray}
	f(t) & = & \left\{\begin{array}{ll}
		1-\lambda\left[z(t)/R_1,\frac{R_2}{R_1}, u_1, v_1\right]/\left(1+\frac{F_2}{F_1}\right) & \text{Primary}\\
		1-\lambda\left[z(t)/R_2,\frac{R_1}{R_2}, u_2, v_2  \right]/\left(1+\frac{F_1}{F_2}\right) & \text{Secondary} 
	\end{array} \right.
\end{eqnarray}
where $z(t)$ is the sky-projected separation of the centers of the two stellar components and $\lambda$ is the fraction of the eclipsed disk blocked by the occulter, given analytically by Mandel \& Agol (2002). The limb darkening coefficients $u$ and $v$ parameterize the radial brightness profile, $I(r)$,  of either binary component as
\begin{eqnarray}
	\frac{I(r)}{I(0)} & =& 1-u \left(1-\sqrt{1-r^2}\right)-v  \left(1-\sqrt{1-r^2}\right)^2
\end{eqnarray}
where $r$ is the projected radial distance from the stellar center, normalized to $R_\star$

The out-of-eclipse light curve was modeled as the product of a harmonic series, $h(t)$, and a low order, quarter specific polynomial, $p_{Q}(t)$ (with $Q = 0,1$); 
\begin{eqnarray}
	h(t) & = & 1+ \sum^{3}_{k = 1} A_k \sin\left[k \phi(t)\right]+ \sum^{3}_{k = 1} B_k \cos\left[k \phi(t)\right] \\
	p_Q(t) & = & \sum^{N_{Q}}_{n= 0} C^{(Q)}_n \left(t-t_0\right)^n  
\end{eqnarray}
for a set of amplitudes  ($A_k$, $B_k$) and polynomial coefficients [$C^{(Q)}_n$] where $\phi(t) = (2\pi/P) (t-t_{\rm ic})$, and $t_0$ is some fixed time near the start of Q0, $N_{0} = 3$ and $N_{1} = 6$.  

The full light curve model, $\ell(t)$, is given as the product of all three variations, i.e., \strike{$\ell(t) = f(t) h(t) p^{Q}(t)$} \ron{$\ell(t) = f(t) h(t) p_{Q}(t)$}.  

We fitted the full model to the data with $R_2/R_1$, $a/R_1$, $F_2/F_1$, $u_1$, $u_2$, $P$, $i$, $e$, $\varpi$, $t_{\rm ic}$, $A_k$, $B_k$ and $C^{(Q)}_n$ as free parameters.  We fixed the quadratic limb darkening coefficient of the primary to $v_1 = 0.2964$, as would be expected (according to Sing 2010) for a star observed in the {\em Kepler} bandpass and having the stellar parameters $T_{\rm eff}=10000$ K, $\log{g} = 4.0$ and $[M/H]=-0.10$ as determined for KHWD3 and tabulated in the  {\em Kepler} Input Catalog. We adopted a linear limb darkening model for the secondary by fixing the quadratic term $v_2$ to zero; the inclusion of this degree of freedom had little effect on the results but was added for completeness. 

The continuously defined model, $\ell(t)$, was numerically integrated before being compared with the long cadence {\em Kepler} light curve.  In detail, for each measured time, $t_j$, we take $n_j$ uniform samples $t_{j,k}= t_j+k \Delta t_j-\tau_{\rm int}/2$, separated by $\Delta t_j = \tau_{\rm int}/n_j$, over the long cadence integration interval of $\tau_{\rm int} = 30$ minutes.  Then, these times were evaluated with $\ell(t)$ and averaged to determine the integrated model flux, $\mathcal{F}_j$, at time $t_j$:
\begin{eqnarray}
	{\mathcal F}_j = \left(\frac{1}{n_j}\right)\sum^{n_j}_{k=0} \ell(t_{j,k}).
\end{eqnarray}
The required number of samples, $n_j$, depends on the curvature of the model relative to the photometric precision at a given time; $n_j > 1$ was only necessary during the relatively sharp eclipse and transit events.  In practice, we determined $n_j$ by increasing its value until the goodness-of-fit statistic (defined below) plateaued.  We settled upon $n_j=60$ for any $t_j$ within $\sim2$ hr of an eclipse or transit and $n_j=1$ elsewhere.

We determined the best fit model to the data by minimizing the $\chi^2$ goodness-of-fit statistic, defined as
\begin{eqnarray}
	\chi^2 = \sum_{s} \frac{\left({\mathcal F}_s-F_s\right)^2}{\sigma^2} \label{eq:chi2}
\end{eqnarray}
where $F_s$ is the measured flux at time $t_s$ and $\sigma$ is the expected statistical error in the flux measurements.  We selected $\sigma = 130$~ppm such that the reduced $\chi^2$ was unity for the best fit solution.

\begin{figure*}[htbp] 
  \epsscale{1.}
  \plotone{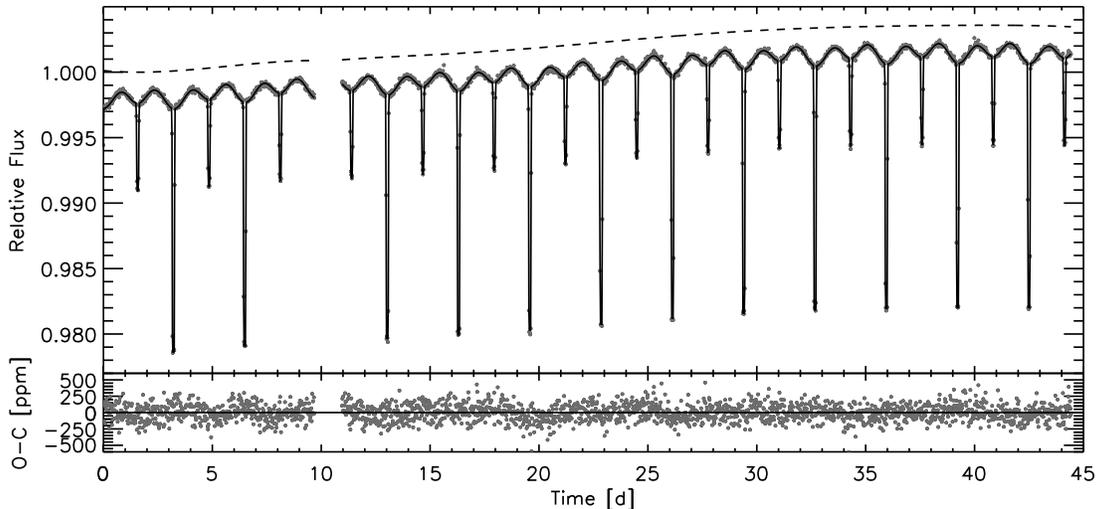}
   \caption{{\em Kepler} light curve for KIC 10657664.  The solid line gives the best fit solution to the data for the model given in \S 2. The dashed curve is the estimated systematic component of the full model. The lower panel shows residuals from subtracting the best fit model from the data (in units of parts per million).    }
   \label{fig:lc}
\end{figure*}

\begin{figure*}[htbp] 
  \epsscale{1.}
  \plotone{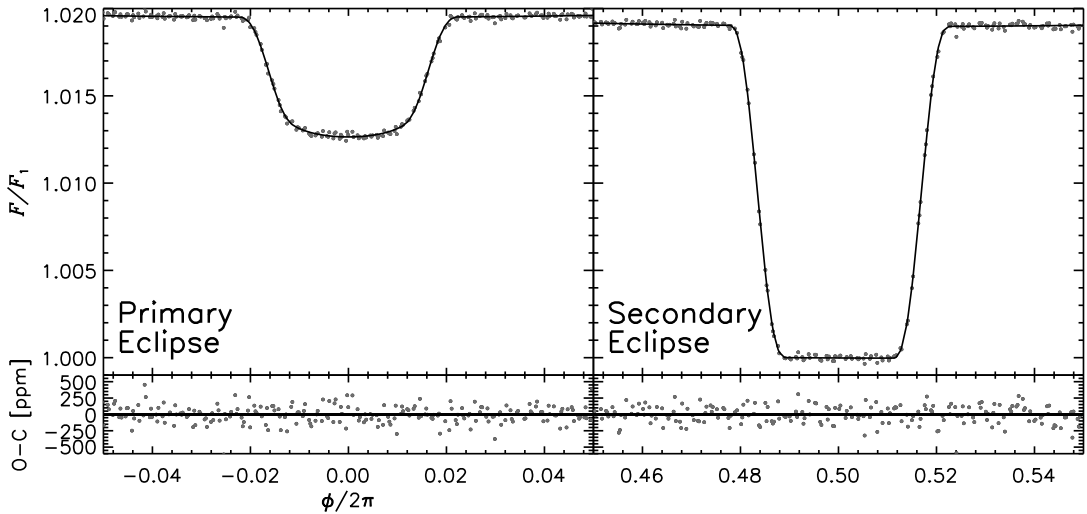}
   \caption{Period-folded, systematic corrected and normalized {\em Kepler} light curve for KIC 10657664 near the phases of eclipse and transit.  The solid lines give the best fit solution for the model described in \S 2.  The lower panel shows residuals from subtracting the best fit model from the data.  }
   \label{fig:elc}
\end{figure*}

\begin{figure*}[htbp] 
   \plotone{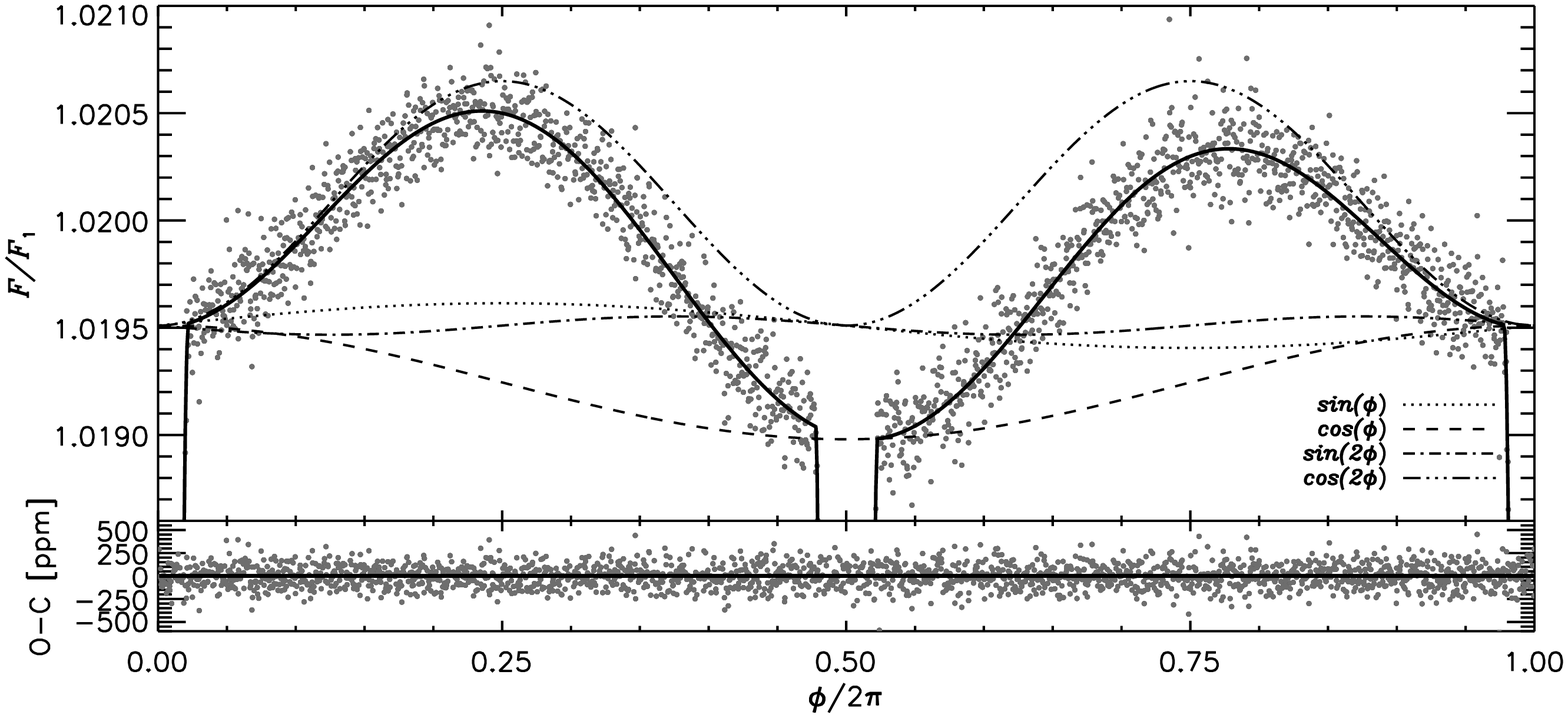}
   \caption{Period-folded, systematic-corrected and normalized {\em Kepler} light curve for KIC 10657664 showing the harmonic content.  The solid lines give the best fit solution to the data for the model given in \S 2.  The four highest amplitude harmonics are plotted individually.  }
   \label{fig:var}
\end{figure*}

The solid curves in Figures~\ref{fig:lc},~\ref{fig:elc}, and \ref{fig:var} trace the best fit solutions to the data. 

\subsection{Markov chain Monte Carlo Analysis}

We determined the posterior probability distributions for the fitted parameters by employing a Markov chain Monte Carlo (MCMC) algorithm.  With the MCMC algorithm, first, a randomly chosen parameter is perturbed by an amount drawn from a normal distribution of some fixed width.  Next, this new parameter and the remaining unperturbed parameters are used to compute $\chi^2$ [as defined in Eqn.~(\ref{eq:chi2})] to determine a likelihood $\mathcal{L'} \propto \exp(-\chi^2/2)$.  The new parameter is ``accepted'' according to a Metropolis-Hasting jump condition: If the likelihood $\mathcal{L'}$ is greater than the unperturbed likelihood $\mathcal{L}$ then the perturbation is accepted, otherwise it is accepted with a probability of $\mathcal{L'}/\mathcal{L}$.  After many perturbations, the resulting ``chain'' of accepted parameters samples the desired joint posterior probability distribution.  For a more detailed introduction to the MCMC algorithm, refer to the appendix of the work by Tegmark et al. (2004).

We generated a chain of $2\times10^6$ links, having selected perturbation widths such that $\sim40\%$ of jumps were accepted.  The chain was checked for adequate mixing and convergence by visual inspection, and by observing that the number of links was much larger than the autocorrelation length (equal to the number of links at which the chain autocorrelation drops below one half) for any selected parameter.  We report, in Table~\ref{tab:phot}, the $50\%$, $15.8\%$ and $84.2\%$ values \ron{(corresponding to the median and $\pm1$--sigma values)} of the cumulative distribution for each parameter, marginalizing over the remaining parameters.  In the upper-left corner of Figure~\ref{fig:dist}, we show the joint distributions for the parameters $R_2/R_1$, $F_2/F_1$, $a/R_1$, $i$ and $u_1$ to demonstrate their correlations.   The remaining parameters, including the harmonic function amplitudes, were weakly correlated with other parameters. 

\subsubsection{Out-of-eclipse harmonic content}

We measured amplitudes (being significantly different from zero) for, in order of decreasing amplitude, the $\cos(2 \phi)$, $\cos(\phi)$, $\sin(\phi)$, $\sin(2 \phi)$ and $\cos(3 \phi)$ modes of the out-of-eclipse variation.  Repeating the above analysis including only  an integral number of orbital cycles did not modify these results.  The four largest modes are plotted individually for comparison in Figure~\ref{fig:var}. The fitted values of the 6 harmonic amplitudes are summarized in Table 1.

In \S 3, we associate these modes with three possible physical effects and describe an analysis used to determine additional system parameters  (such as the mass ratio $q$), assuming the validity of these associations. 

{\em Correlated Noise} -- We have assumed a ``white'' noise spectrum in both our best-fit and MCMC analyses.  However, we note that correlated noise could have a significant affect on our estimates of harmonic content.  To investigate this possibility further, we completed a secondary analysis, as follows, whereby we determined what we consider to be the most conservative errors on these amplitudes.  

We first corrected for the low-frequency trend by dividing the data by the best-fit estimate for $p_Q(t)$.  Then, having fixed the best-fit estimates for the eclipse parameters, we fit a unique harmonic model to each individual cycle (i.e., by using only the data from one primary transit to the next).  We then determined the mean amplitude, for each mode, and the standard deviation in that mean averaging over the 12 available cycles.   

We found that the scatter across all cycles, in a given harmonic amplitude, was consistent with random noise (i.e., no minority of cycles was significantly biasing the mean). The means across all modes agree with the MCMC-determined amplitude medians within the MCMC-determined errors.  The standard deviations in the means were larger, as may be expected in the presence of correlated noise, than the amplitude uncertainties determined via MCMC.  However, only the $\sin \phi$ and $\cos \phi$ modes have substantially higher uncertainties, with each being roughly twice as large as the MCMC estimates.  These inflated, and likely overestimated, errors translate into a $\sim 10\%$ correction in the {\em uncertainty} in the secondary mass estimate (as determined in \S3).  All other system parameters were less significantly affected.  We therefore decided to use the MCMC error estimates in the subsequent analysis.

\subsubsection{Eccentricity}

We also report a weak detection of eccentricity for this system, having measured a non-zero value for \strike{$e \cos(\varpi) = 0.003\pm0.0005$}\ron{$e \cos(\varpi) = 0.0029\pm0.0005$} as inferred from the $\sim 4.5$ minute delay (relative to the circular orbit expectation) of the start of secondary eclipse.  We note that we did not account for delays as a result of the finite speed of light crossing the orbit;  however, given the estimated separation of the binary (see \S 3), we expect light travel time delays of $\sim 30$ seconds.  This correction corresponds to an error in $e \cos\varpi $ that is smaller than the estimated statistical error.  In \S 3.6.1, we comment on the consequences of a non-zero eccentricity in the interpretation of the harmonic modes of the out-of-eclipse light curve.

\begin{deluxetable}{lcll}
\tablecolumns{4}
\tablewidth{0pt}
\tablecaption{KHWD3 Light Curve Model Parameters}
\tablehead{Parameter & Median &  84.2\% \ron{(+1$\sigma$)}& 15.8\% \ron{(-1$\sigma$)}} \\
\startdata
$P$ [day] &3.273713 & +0.000008 & -0.000008\\
$t_{\rm ic}$ [BJD] & 2454951.85857& +0.00012 &-0.00012\\  
$e\cos(\varpi)$ & 0.0029 & +0.0005 & -0.0005\\ 
$|e\sin(\varpi)|$ & \multicolumn{3}{l}{$\lesssim 0.1^{\rm a}$}\\ \\
\multicolumn{4}{l}{Eclipse Parameters} \\ \hline
$R_2/R_1$ $[\times10^2]$ &8.100&+0.024&-0.020\\
$F_2/F_1$ $[\times10^2]$ &1.900&+0.002 &-0.002\\
$a/R_1$ &7.02 & +0.08 & -0.10\\
$i$ [deg] &84.46 &+0.14 &-0.18\\
$u_1$ &0.20 & +0.02 &-0.02\\
$u_2$ & \multicolumn{3}{l}{unconstrained$^{\rm b}$} \\  \\
\multicolumn{4}{l}{Out-of-eclipse Harmonic Amplitudes [$\times10^4$]} \\ \hline
$A_1 [\sin(\phi)]$ &  1.03&+0.04 &-0.04\\
$B_1 [\cos(\phi)]$ &  2.64&+0.04&-0.04\\
$A_2 [\sin(2\phi)]$ & -0.44&+0.04&-0.04\\
$B_2 [\cos(2\phi)]$ & -5.68&+0.04&-0.04\\
$A_3 [\sin(3\phi)]$ &  0.02&+0.04&-0.04\\
$B_3 [\cos(3\phi)]$ & -0.13&+0.05&-0.05
\enddata
\tablenotetext{a}{$e \sin \varpi$ is nearly unconstrained by the data (and uncorrelated with the remaining parameters).  Here we report the full range of its weakly converged Markov chain.  }
\tablenotetext{b}{The limb darkening parameter of the secondary was limited to positive values less than $0.8$.  }
\label{tab:phot}
\end{deluxetable}

\section{Light Curve Model-Independent System Parameters}

From the measured values of $R_2/R_1$ and $F_2/F_1$ we may determine the mean surface brightness ratio $\mathcal{B}_2/\mathcal{B}_1$ between the secondary and primary according to
\begin{eqnarray}
	\frac{F_2}{F_1} \left(\frac{R_1}{R_2}\right)^2 & = & \mathcal{B}_2/\mathcal{B}_1 \nonumber \\
		&\simeq & \frac{\exp(hc/\lambda k T_1) -1}{\exp(hc/\lambda k T_2) -1}
\end{eqnarray}
where in the second line we have substituted for the surface brightness ratio the expectation assuming black-body emission and a narrow observed wavelength range (with $\lambda \approx 6000$\AA~for {\em Kepler}).  \ron{To determine this brightness ratio in practice, for a given set of temperatures, we integrated each black-body spectrum over the wide {\em Kepler} response function.}

The harmonic content of the out-of-eclipse light curve variation for KHWD3 is presumed to be astrophysical, i.e., related to the binary system, rather than systematics.  In the previous section, we reported on the measured values of the harmonic amplitudes, $A_k$ and $B_k$ for $k = 1$, $2$, or $3$, where we remind the reader that the out of eclipse variation was modeled as
\begin{eqnarray}
	h(t) & = & 1+ \sum^{3}_{k = 1} A_k \sin\left[k \phi(t)\right]+ \sum^{3}_{k = 1} B_k \cos\left[k \phi(t)\right].
\end{eqnarray}
	
Briefly, we expect that the $\sin(\phi)$ term is due to Doppler boosting (hereafter ``DB''), the $\cos(\phi)$ term is largely due to mutual illumination (hereafter ``ILL'') between the two components, and the $\cos(2\phi)$ and $\cos(3\phi)$ terms are largely due to the ellipsoidal shape deformation of the primary by the secondary's tidal field (so-called ellipsoidal light variation, hereafter ``ELV''). In this nominal model, we do not expect non-zero amplitudes for the $\sin(2 \phi)$ or $\sin(3 \phi)$ modes.

\subsection{Expected harmonic amplitudes}
\label{sec:harmonic}

We assume a circular binary orbit throughout the following discussion.  A departure from this assumption is discussed in \S 3.6.1. 

The predicted amplitudes $A_k$, $B_k$ may be decomposed into a sum of effect-specific amplitudes such that
\begin{eqnarray}
	A_k & = & A^{\rm DB}_k + A^{\rm ELV}_k+A^{\rm ILL}_k \\
	B_k & = & B^{\rm DB}_k + B^{\rm ELV}_k+B^{\rm ILL}_k .
\end{eqnarray}
In turn, the effect-specific amplitudes are a sum of the individual contributions from each member of the binary; for example, $A^{\rm DB}_k = A^{\rm DB, 1}_k+A^{\rm DB, 2}_k$. 

{\em Doppler boosting} -- The \strike{spectrum}\ron{apparent spectral intensity} of either component changes throughout the orbit as a result of Doppler shifting, photon emission rate modulation and beaming.  The combination of these effects is called Doppler boosting; it is discussed in the context of {\em Kepler} by Loeb \& Gaudi (2003) \ron{and by Zucker et al. (2007)} and was detected for at least one of the two other candidate hot white dwarf companions in the {\em Kepler} field (KOI-74 and KOI-81; van Kerkwijk et al.\,2010) \ron{and for one planetary system (CoRoT-3; Mazeh 
\& Faigler, 2010)}.  The time variability of the Doppler boosting signal is carried by the projected radial velocity $v_{r,(1,2)}$ where, for a circular orbit, $v_{r,(1,2)} \propto \sin \phi$ and the total contribution to the harmonic out-of-eclipse content by Doppler boosting is\strike{, to good approximation,} 
\begin{eqnarray}
	A^{\rm DB}_{1} & \ron{=} & \alpha_1 \left(\frac{v_{r,1}}{c}\right)+\alpha_2 \left(\frac{v_{r,2}}{c}\right)  \nonumber  \\
		& = & \left(\frac{2 \pi a}{P c}\right)\left[ \alpha_1\left(1+\frac{1}{q}\right)^{-1} -\alpha_2  \left(1+q\right)^{-1}\ron{\frac{F_2}{F_1} }\right]\ron{.}
\end{eqnarray}
\strike{with}
\strike{where $\lambda \approx 6000$\AA ~corresponds to the center of the wide {\em Kepler} response function.  The above assumes a black-body model for the emission and a uniform stellar brightness profile. If one evaluates a proper average of equation (12) over the Kepler band pass, and/or includes a limb darkening profile for the star, the coefficient $\alpha$ changes by only a few percent.}

\ron{The DB prefactor ($\alpha$) depends on the shape of the emission spectrum in the emitter's rest frame such that}
\ron{\begin{eqnarray}
	\alpha & \simeq & 3-\langle \frac{d \ln F_\nu}{d \ln \nu} \rangle
\end{eqnarray}}
\ron{where the average is weighted over the {\em Kepler} response.  It has been empirically validated in the case of KOI-74 \ron{(Ehrenreich et al.\,2010)} that prefactors estimated assuming a black-body emission model are overestimated by greater than 20\% for stars with $T \approx 10000$ K.  As such, we utilize the synthetic spectra by Kurucz (\ron{1979}) to generate DB prefactors over a range of effective temperatures, by fitting power law models \ron{(an impressively good approximation over the Kepler band pass)}.  The fits are performed in a weighted least-squares sense with the weight at a given frequency being specified by the {\em Kepler} response function.  For the A star primary, we selected model spectra of fixed gravity ($\log \,g = 4.0$ [cgs]) over a range of temperatures from 8250 K to 11500 K and derived DB prefactors.  When then fitted a \ron{linear} function in temperature to these prefactors for use in our subsequent analyses:} \ron{
\begin{eqnarray}
	\alpha_1(T_1) & \approx & 1.90-1.12 \left(\frac{T_1}{10000 {\rm K}}-1\right) 
\end{eqnarray} }

\ron{The DB prefactors for the WD component [$\alpha_2(T_2)$] were computed analogously\bon{, but  assuming a simple} black-body emission model. } 

\vspace{0.2cm}
{\em Ellipsoidal light variation} -- The mutual gravitational interaction between members of a binary will induce nominally prolate distortions of their surfaces having major axes along the line connecting their centers.  As a result, the sky-projected cross-sectional area, and consequently the observed flux, are time variable (Morris 1985; see also Pfahl, Arras, \& Paxton 2008).  For a circular orbit, the expected maxima of this variation occur twice per orbit at the quadratures [generating a large $\cos(2 \phi)$ amplitude].  Kopal (1959) calculated the ELV contribution to the out-of-eclipse harmonic content as (where we follow the notation by Morris 1985):
\begin{eqnarray}%
	B^{\rm ELV}_1  & \simeq &  -3 Z_1(3) q  \left(\frac{R_1}{a}\right)^4 \sin i  \\
	B^{\rm ELV}_2  & \simeq & -Z_1(2) q  \left(\frac{R_1}{a}\right)^3 \sin^2 i  \\
	B^{\rm ELV}_3  & \simeq & -5 Z_1(3) q  \left(\frac{R_1}{a}\right)^4 \sin^3 i 
\end{eqnarray} 
where the prefactors $Z_1(2)$ and $Z_1(3)$ (Eqn.~3 in the work by Morris 1985) depend on limb-darkening and gravity darkening parameters with $Z_1(2) \approx 1.5$ and $Z_1(3)/Z_1(2) \approx 0.05$.  In our calculations, we assume a fixed linear limb-darkening parameter [$u = 0.44$; according to Sing (2009) for the {\em Kepler} Input Catalog stellar parameters listed in \S2.1] and a gravity darkening parameter that depends on temperature according to the von Zeipel law (Eqn.~10 in the work by Morris 1985). 

The contribution to the ELV due to the distortion of the secondary by the primary is smaller by a factor of at least $(R_2/R_1)^3 (F_2/F_1) /q^2 \sim 0.05\%$ and is therefore safely ignored.

{\em Mutual illumination} -- It is expected that radiation arising from one binary component and incident on the other will subsequently either be scattered or absorbed and reemitted.  As such, the surfaces of the components that are directly facing one another will be brighter than other regions.  As the binary orbits, we perceive these phases as a modulation of the out-of-eclipse total flux level.  The maximum illumination of the primary is at inferior conjunction while that of the secondary is at superior conjunction, contributing with opposite signs to a predominately $\cos \phi$ mode of the out-of-eclipse variation. If the binary is in radiative equilibrium, then all incident radiation (bolometric) must be absorbed and reemitted at approximately  the effective temperature of the illuminated member (Eddington 1926).  In this scenario, Kopal (1959) again provides the expected out-of-eclipse harmonic content due to mutual illumination as
  \begin{eqnarray}
  	B^{ILL}_1 & \simeq & \frac{17}{16} \left(\frac{R_1}{a}\right)^2 \left[ \frac{1}{3}+\frac{1}{4}\left(\frac{R_1}{a}\right)\right] \left(\frac{T_2}{T_1}\right)^4 \left(\frac{R_2}{R_1}\right)^2- \nonumber\\
	 & & \frac{17}{16} \left(\frac{R_2}{a}\right)^2 \left[ \frac{1}{3}+\frac{1}{4}\left(\frac{R_2}{a}\right)\right] \frac{BC_2}{BC_1} \\
	 B^{ILL}_2 & \simeq & \frac{17}{16} \left(\frac{R_1}{a}\right)^2 \left[ \frac{16}{27 \pi^2}+\frac{3}{16}\left(\frac{R_1}{a}\right)\right] \left(\frac{T_2}{T_1}\right)^4 \left(\frac{R_2}{R_1}\right)^2+ \nonumber\\
	 & & \frac{17}{16} \left(\frac{R_2}{a}\right)^2 \left[ \frac{16}{27 \pi^2}+\frac{3}{16}\left(\frac{R_2}{a}\right)\right]  \frac{BC_2}{BC_1}
\end{eqnarray}
where $BC$ is an approximate bolometric correction to the visual (see Allen 1964) with 
\begin{eqnarray}
	-2.5 \log BC & = & -42.5+10 \log T_{\rm eff}+\frac{29,000 {\rm K}}{T_{\rm eff}}.
\end{eqnarray}

\subsection{First cut at determining the constituent masses: an inconsistency}

In principle,  we may solve for the system parameters $M_1$, $T_1$, $R_1$, $M_2$, $T_2$, $R_2$ and $a$ using the formalism introduced in the previous sections (Eqns.\,7, 11-19), Kepler's third law, and the results from the light curve analysis presented in \S 2.  However, the illumination effect (given by Eqns.\,17 and 18) is nearly independent of the system masses, which are largely determined by the Doppler boosting (Eq.\,11) and ELV effects (Eqns.\,14-16).  If we take the simplest forms of both the DB and ELV effects we find:
\begin{eqnarray}
A^{\rm DB}_1 & \simeq &\ron{2.0} \left(\frac{2 \pi a}{P c}\right) \left(\frac{q}{1+q}\right) \\
B^{\rm ELV}_2  & \simeq & -1.5  q  \left(\frac{R_1}{a}\right)^3 
\end{eqnarray} 
where the coefficients $\alpha_1$ and $Z_1(2)$ have been set equal to their nominal values, $i$ has been set to $90^\circ$ without loss of accuracy, and $q \equiv M_2/M_1$.  For purpose of this particular exercise, we have neglected the relatively small effect of the Doppler boosting of the hot white dwarf.  If we eliminate $a$ in favor of $P$ through Kepler's third law, and plug in our measured values for $P$ and $R_1/a$, we find:
\begin{eqnarray}
A^{\rm DB}_1 & \simeq &  \ron{9.6 \times 10^{-4}} m_1^{1/3}  \frac{q}{(1+q)^{2/3}} \simeq 1.03 \times 10^{-4}\\
B^{\rm ELV}_2  & \simeq & -4.3 \times 10^{-3}  q \simeq -5.68 \times 10^{-4}
\end{eqnarray} 
where the numerical values on the right sides of the equations are taken from Table 1, and the mass of the primary, $m_1$ has been expressed in solar units.  Since these equations depend on $q$ and only very weakly on the actual mass of the primary, they both, in effect, independently determine $q$.  For any primary mass between 2 and 3 $M_\odot$, the two equations yield: \strike{$q \simeq 0.064 \pm 0.004$}\ron{$q \simeq 0.078$--$0.090$} and $q \simeq 0.13$, respectively, depending on whether we utilize the DB or ELV amplitudes to determine $q$.  These are clearly inconsistent, and this remains the case, even when we later utilize all the terms in Eqns.\,(7), (11)--(19), and take full account of the statistical uncertainties associated with the measure quantities.  Since there is a dependence on $M_1$ in one of the two equations, there exists a formal solution to the two equations for $M_1$ and $M_2$ (\strike{0.15 $M_\odot$ and 0.02 $M_\odot$}\ron{0.7 $M_\odot$ and 0.1 $M_\odot$}, respectively).  However, this solution is not physically plausible, and therefore we are left with this intriguing inconsistency in the value of $q$ that we determine.

\subsection{Roughly determined parameters: $T_1$ and $\rho_1$ }

Once we decide on the mass ratio for the system, we will need an estimate of the mass of the primary in order to determine the mass of the white dwarf.  There are two accessible parameters that we can utilize to estimate the primary mass: the effective temperature, $T_1$, and the mean stellar density, $\rho_1$.  $T_1$ is listed for the primary star in the {\em Kepler} Input Catalog (KIC) as 10,500 K, based on 5-color photometry.  We have no readily available estimate of the uncertainty in the KIC value for $T_1$, but we estimate that it could be $\sim \pm 500$ K.  As we shall see in \S3.5, the illumination effect can be used to directly infer the effective temperature of the primary, and yields a result of \strike{ $9100 \pm 150$}\ron{$9500 \pm 150$} K, which would then be some \strike{2.7}\ron{2} $\sigma$ away from the KIC tabulated value.   We therefore consider a range for $T_1$ that includes both the KIC value and the one we derive from the illumination effect.

The second readily, and more accurately, determined parameter associated with the primary is its mean stellar density, $\rho_1 \equiv M_1/(4/3 \pi R_1^3)$.  We point out that $\rho_1$ is only a function of the light curve-determined parameters $P$ and $a/R_1$, with a weak dependence on $q$.  This dependence of $\rho_1$ on these parameters may be found by dividing Kepler's third law by $R_1^3$:
\begin{eqnarray}
	\left(\frac{a}{R_1}\right)^3 & = & \frac{1}{3 \pi} G P^2 \rho_1 (1+q)
\end{eqnarray}
This relation was pointed out by Seager \& Mall{\'e}n-Ornelas (2003) in the context of transiting exoplanets.   With exoplanets, the value of $q$ is much smaller than the error in $a/R_1$ such that $\rho_1$ can be determined independently of any mass information.  While the same is not quite true with our binary, the value of $q$ is still small enough (with $q \lesssim 0.2$ as a very conservative estimate) such that a moderately accurate value of $\rho_1$ may be determined independently of $q$.  In particular, we find that $\rho_1 \approx 0.61\pm0.02$ g cm$^{-3}$ by using Eqn.~(24) with $q=0$ and the results from our photometric analysis.  We consider this estimate to be maximal and accurate to within $20\%$ of its true value.

\subsection{A prior on $M_1$}

In the previous section, we argued that $T_1 \sim$ 10,000 K and that $\rho_1 \lesssim$ 0.61 g cm$^{-3}$.  Given these two values, we can motivate a prior on $M_1$.  

If we assume the primary is on the zero age main sequence (ZAMS), then we may utilize the approximate formula by Eggleton (2006) to estimate its mass as a function of either effective temperature or density.  Picking the former to be in the range 9000 K $\lesssim T_1 \lesssim$ 11,000 K, we find a mass 2.0 $\lesssim M_1/M_\odot$ $\lesssim$ 2.6.  Choosing the latter to be in the range 0.51 g cm$^{-3}$ $\lesssim \rho_1 \lesssim$ 0.61 g cm$^{-3}$, we find a mass 2.5$\lesssim M_1/M_\odot\lesssim 3.2$.  These ranges in $M_1$ intersect, suggesting $M_1\sim2.5$ $M_\odot$; alternatively, either the primary has aged somewhat off the ZAMS or the metallicity of the primary is different from Solar.  To investigate these possibilities further, we utilized the Yonsei-Yale stellar evolution isochrones (Yi et al.\,2001) to determine a likely mass range subject to the priors on temperature and density as given above and additionally a normally distributed prior on the metallicity, [Fe/H] = 0.0$\pm0.2$ (applying the methodology described by Carter et al.\,2009).  From this analysis, we find that $M_1 = 2.3^{+0.5}_{-0.4}$ $M_\odot$ with an estimated age for the star in the range 200--600 Myr; although, given a likely mass transfer history (see \S 5), the interpretation of this age estimate is unclear.

Considering the above ranges in $M_1$, we opted for a conservative mass prior for the primary being normally distributed about $M_1 = 2.5$ $M_\odot$ with an rms width $0.5$ $M_\odot$.

\subsection{Derived system parameters subject to the prior on $M_1$}

In the following analysis of the system parameters we assume that the illumination model is reasonably correct (Eqns.\,17 and 18), and we utilize {\em either} the ELV effect (Eqns.\,14-16) {\em or} the Doppler boosting effect (Eqns.\,11 and 12).
Given the common prior on $M_1$ (see the previous section), either effect (ELV or DB)
results in an independent estimate of $q$.  As discussed earlier, however, we find that these estimates disagree with one another: ignoring DB [by nullifying the statistical weight of the measurement of the $\sin(\phi)$ mode] gives $q \simeq 0.15\pm0.01$, whereas ignoring ELV [by nullifying the statistical weight of the measurement of the $\cos(2\phi)$ mode] gives \strike{$q \simeq 0.08\pm0.01$}\ron{$q \simeq 0.11\pm0.01$}.   Given this discrepancy, we opted to determine and report system parameters for these two scenarios {\em separately}.  

To determine posterior distributions for all system parameters, we executed another Markov chain Monte Carlo algorithm (as described in \S 2.2) subject to the likelihood $\mathcal{L} \propto \exp(-\chi_T^2/2)$ where
\begin{eqnarray}
	\chi_T^2 &=& \frac{ \left(\Delta\frac{R_2}{R_1}\right)^2}{\sigma^2_{R_2/R_1}}+\frac{ \left(\Delta\frac{F_2}{F_1}\right)^2}{\sigma^2_{{F_2/F_1}}}+\frac{ \left(\Delta\frac{a}{R_1}\right)^2}{\sigma^2_{{a/R_1}}}+\frac{ \left(\Delta i\right)^2}{\sigma^2_{{i}}}+\frac{ \left(\Delta P\right)^2}{\sigma^2_{{P}}}+\nonumber \\
		&& \rho_{R_2/R_1, a/R_1}\frac{ \left(\Delta\frac{R_2}{R_1}\right)\left(\Delta\frac{a}{R_1}\right)}{\sigma_{R_2/R_1}\sigma_{a/R_1}}+\rho_{R_2/R_1, i}\frac{ \left(\Delta\frac{R_2}{R_1}\right)\left(\Delta i\right)}{\sigma_{R_2/R_1}\sigma_{i}}+ \nonumber \\
		&& \rho_{a/R_1, i}\frac{ \left(\Delta\frac{a}{R_1}\right)\left(\Delta i\right)}{\sigma_{a/R_1}\sigma_{i}}+\sum_k \frac{\left(\Delta A_k\right)^2}{\sigma^2_{A_k}}+\sum_k \frac{\left(\Delta B_k\right)^2}{\sigma^2_{B_k}}+\nonumber \\
		&& \frac{ \left(M_1-2.5 M_\odot\right)^2}{\left(0.5 M_\odot\right)^2}
\end{eqnarray}
where for any parameter ``$x$'', $\Delta x$ is the difference between the median estimate of the parameter $x$ and its current value (at a given link in the Markov chain), $\sigma^2_x$ is the variance in the variable $x$ and $\rho_{x,y}$ is the correlation coefficient between variables $x$ and $y$.  The current values of the harmonic amplitudes $A_k$ and $B_k$ are determined as a function of the current values of the parameters $M_1$, $M_2$, $T_1$, $T_2$, $R_1$, $R_2$ using Eqns.~(7), (11)--(19) and Kepler's third law. The medians, variances and correlation coefficients are fixed to their values as calculated using the joint posterior distribution resulting from the earlier analysis of the light curve data (as described in \S 2). 

To ignore, in effect, the statistical influence of DB or ELV, we remove the terms in $\chi_T^2$  associated with $A_1$ [$\sin(\phi)$] or $B_2$ [$\cos(2 \phi)$], respectively.  In each scenario, a Markov chain of length $5\times10^6$ was generated and the resulting posterior distributions were inspected to have adequately converged.  

\begin{figure*}[htbp] 
\epsscale{1.2}
   \plotone{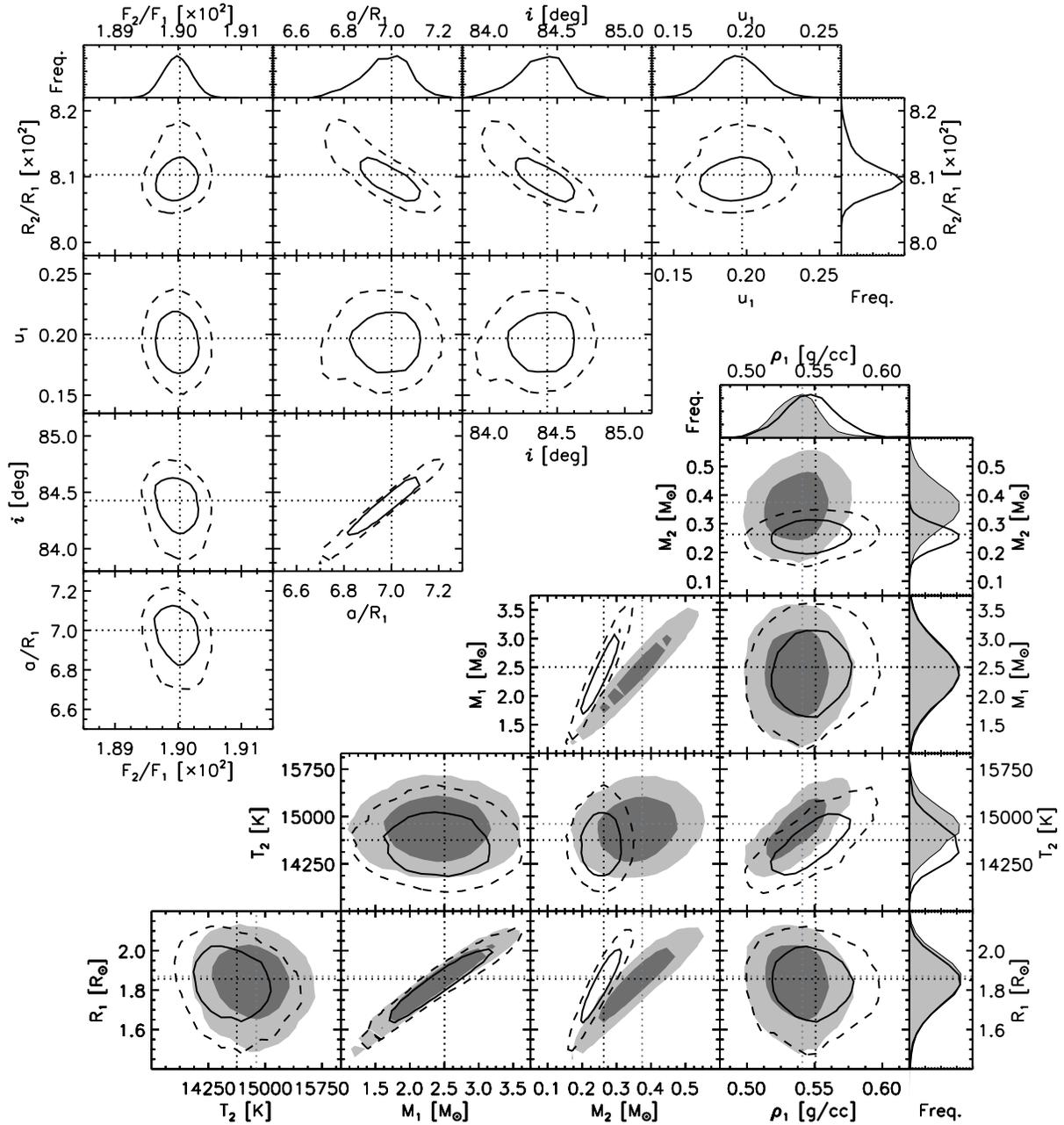}
   \caption{Joint posterior distributions (showing 68\% and 95\% confidence contours) and histograms for KHWD3 light curve model and system parameters.  The upper left portion shows some results from the analysis, as described in \S 2, of the {\em Kepler} light curve data for KHWD3.  See Table~1 for complete results.  The lower right portion shows some results from a subsequent analysis, as described in \S 3, to determine absolute system parameters for KHWD3 given the results from the photometric analysis.  The shaded contours and histograms correspond to the scenario in which only ellipsoidal light variation and illumination were assessed.  The non-shaded contours correspond to the scenario in which only the Doppler boosting and illumination were assessed.  See \S 3 for more details.      }
   \label{fig:dist}
\end{figure*}

We report the results of our analysis of the system parameters in Table~\ref{tab:sys}.  The two columns in the table are the system parameters determined from the ELV and ILL effects, on the one hand, and the DB and ILL effects on the other.  We list the median values of the parameters, as well as the 15.4\% and 84.2\% \ron{(corresponding to the $\pm1$-sigma values)} values of the cumulative distribution function for each parameter of interest, having marginalized over the remaining parameters.  In the lower-right corner of Fig.\,\ref{fig:dist} we show the joint posterior distributions in a number of parameters to show correlations and differences between emphasizing DB or ELV.

Of the approximately dozen system parameters reported in Table 2 (luminosities, radii, $T_{\rm eff}$, masses, gravity, and so forth), the agreement between the two columns (ELV and DB dominated analysis) is remarkably good except for those parameters that involve the white dwarf mass.  The latter depends heavily on the mass ratio which, in turn, is strongly dependent on the assumption of whether the ELV or DB amplitudes correctly reflect their respective physical quantities.

The estimated masses, temperatures and radii of the secondary, in either of the two scenarios, are consistent with a hot white dwarf companion.  In \S 5, we discuss possible evolutionary histories leading to these two outcomes and estimate the properties the system progenitors.

\begin{deluxetable}{lll}
\tablecolumns{3}
\tablewidth{0pt}
\tablecaption{KHWD3 Model-Independent System Parameters}
\tablehead{ Parameter & ELV+ILL Model & DB+ILL Model} \\
\startdata
$a$ [AU] & 0.061$\pm$0.004 & 0.061$\pm$0.004 \\
$M_2/M_1$ & 0.150$\pm$0.006&0.105$\pm$0.008\\
$T_2/T_1$ & 1.56$\pm$0.01&1.55$\pm$0.01\\
\\ \hline
$M_1$ $[M_\odot]$ & $2.5\pm0.5^{\rm a}$ & $2.5\pm0.5^{\rm a}$ \\
$R_1$ [$R_\odot$] & 1.87$\pm$0.13&1.86$\pm$0.12\\
$T_1$ [K] & 9600$\pm$150&9500$\pm$150\\
$L_1$ [$L_\odot$] & 26.2$\pm$ 3.8&24.8$\pm$ 3.3\\
$\rho_1$ [g/cc] & 0.54$\pm$0.02&0.55$\pm$0.02\\
$\log(g_1)$ [cgs] & 4.29$\pm$0.03&4.30$\pm$0.03\\
\\ \hline
$M_2$ [$M_\odot$] & 0.37$\pm$0.08&0.26$\pm$0.04\\
$R_2$ [$R_\odot$] & 0.151$\pm$0.011&0.150$\pm$0.010\\
$T_2$ [K] & 14900$\pm$300&14600$\pm$300\\
$L_2$ [$L_\odot$] & 1.00$\pm$0.15&0.93$\pm$0.13\\
$\rho_2$ [g/cc] & 153$\pm$11&109$\pm$9\\
$\log(g_2)$ [cgs] & 5.65$\pm$0.04&5.50$\pm$0.02\\ \\
\multicolumn{3}{l}{Observed-Calculated Harmonic Amplitudes [$\times10^4$]}\\ \hline
$A_1$ [$\sin(\phi)$] & -0.48$\pm$ 0.12$^{\rm b}$& 0.00$\pm$ 0.06\\
$B_1$ [$\cos(\phi)$] &  0.01$\pm$ 0.06& 0.00$\pm$ 0.06\\
$A_2$ [$\sin(2\phi)$] & -0.44$\pm$ 0.04$^{\rm b}$&-0.44$\pm$ 0.04$^{\rm b}$\\
$B_2$ [$\cos(2\phi)$] & -0.00$\pm$ 0.06&-1.80$\pm$ 0.35$^{\rm b}$\\
$A_3$ [$\sin(3\phi)$] &  0.02$\pm$ 0.04& 0.02$\pm$ 0.04\\
$B_3$ [$\cos(3\phi)$] &  0.08$\pm$ 0.05& 0.02$\pm$ 0.05 
\enddata
\label{tab:sys}
\tablecomments{a = Mass of the primary is assumed.   b = Values did not affect the likelihood. See \S~3 for details.}
\end{deluxetable}

\subsection{Possible missing physics}

In the previous sections, we inferred system parameters for KHWD3 only after discounting certain aspects of our model for the out-of-eclipse harmonic content in order to resolve some discrepancies in the measured amplitudes that exceed the estimated statistical uncertainties.  At this time we admittedly have no satisfactory explanation for these discrepancies, though it is comforting that they do not affect any of our system parameter determinations other than those involving the white-dwarf mass -- and even then the white dwarf mass is uncertain by at most a factor of 2.  We note that, to the best of our knowledge, this is the first time that all three effects, DB, ILL, and ELV, have been studied at such small amplitudes (e.g., parts in $10^4$).  However, in this regard, we would have surmised that the simple, classical, analytic models for these effects should work even better at these smaller amplitudes than at larger amplitudes.  It is also possible that there are some additional, heretofore unexplained, physical effects which complement the harmonic content that we have not attempted to model, e.g. the O'Connell effect (see, for example, Davidge 
\& Milone 1984).

\subsubsection{Non-zero eccentricity}

We measured a tentative non-zero value for the eccentricity $e$, at weak significance, with \strike{$e \cos \varpi = 0.003\pm0.0005$}\ron{$e \cos \varpi = 0.0029\pm0.0005$};  however, we assumed a circular orbit in assessing the out-of-eclipse harmonic content.  A small but non-zero eccentricity would lead to slightly modified expectations from all three effects leading to the out-of-eclipse variation (ELV, DB \& ILL).  For very small eccentricity, its influence can be determined analytically with epicyclic approximations to the binary orbit, whereby the semi-major axis $a/R_1$ in the model expressions (Eqns.~11-19) is replaced with the instantaneous separation $r/R_1$, which is a function of $\phi$, $e \cos \varpi$ and $e \sin \varpi$.  We find that DB is the only effect that can have a significantly altered harmonic spectrum [to linear order in eccentricity and given our restriction on $e \cos \varpi$].  This alteration is entirely manifested in the inclusion of a $\sin(2 \phi)$ mode whose amplitude relative to the dominant $\sin\phi$ mode is

\begin{eqnarray}
	\frac{A_2}{A_1} & \approx & e \sin \varpi.
\end{eqnarray}
We did detect a non-zero amplitude $A_2/A_1 = -0.43\pm0.06$, however, this amplitude is much too large to be accounted for by eccentricity alone.

\subsubsection{Departure from synchrony}

We have assumed synchronous rotation between the primary, secondary, and orbit when calculating the contribution to the harmonic content from ELV or ILL (as described in \S 3.1).  Should the primary be rotating at a different rate from the orbital frequency, frictional drag within the primary may force the tidal distortion due to the secondary to lag or lead the orbit (as is the case with the tides imposed on Earth by the Moon). The illumination effect may also lag or lead as a result of both the modified phase variation of the sky-projected cross-sectional area and additionally, depending on the depth in the stellar atmosphere at which radiant energy from the secondary is deposited, from a ``hot spot'' that is shifted from the substellar point.  

To investigate this possibility further, we carried out an analysis where we assumed that both the ELV and ILL effects are shifted by a common variable phase from their nominal circulation.  DB is independent of the assumption of synchrony and was not permitted a phase shift.  Given this additional degree of freedom, we find that we may account for the measured values of all harmonic amplitudes including the relatively large, and previously unexplained, $\sin(2 \phi)$ mode.  However, the results of this analysis remain unphysical 
\strike{($T_1 \approx$ 9,100 K, $M_1 \approx 0.13$ $M_\odot$, $R_1 \approx $ 0.7 $R_\odot$, $T_2 \approx $ 14,000 L, $M_2 \approx $ 0.02 $M_\odot$, $R_2 \approx 0.06$ $R_\odot$)} 
\bon{($M_1 \approx 0.8$ $M_\odot$ and $R_1 \approx $ 1.3 $R_\odot$)}.  
Moreover, we determined that these effects lag the orbit by $5^\circ$ while reasonable expectations for the efficiency of tidal dissipation (quantified via the parameter $Q$) would predict maximal phase offsets of $1/Q\sim10^{-6}$.

\subsubsection{Other effects}

Here, we mention two \,other effects not included in our model that could significantly affect our interpretation of the harmonic content and/or the photometric results:
\begin{itemize}%
	\item Stellar spotting -- A persistent stellar spot on a synchronously rotating primary could induce variations with power in multiple harmonics, most likely in the fundamental mode (competing with DB \& ILL).  The amplitude of this effect depends on the size and flux contrast, relative to the unspotted photosphere, of the spot. While spots in tidally locked binaries may persist for months, it is unlikely that an A star lacking a substantial convective outer envelope would exhibit significant spotting behavior (Strassmeier 2009). \ron{ A possible exception would be a magnetically active A star (e.g., a ``peculiar'' or Ap star; see Kochukhov (2010) and references therein).}
	\item Rapid rotation of the primary -- In the previous section, we discussed possible phase lags in the ELV and ILL effects as a result of non-synchronous rotation.  In addition, the primary may be significantly oblate as a result of very fast rotation.  In this case, $a/R_1$ as inferred in the light curve analysis may be different from the true normalized semi-major axis.  In particular, if the spin axes of the rapidly rotating primary and the orbit are aligned, then $a/R_1$ would likely be overestimated (Barnes 2009).  Additionally, very fast rotation may induce an unexpected stellar brightness profile as a result of strong gravity darkening; both the inferred radius ratio, $R_2/R_1$, and $a/R_1$ may be inaccurate as result (Barnes 2009). However, given the short orbital period of 3.3 days, it is most likely that the orbit and the primary star have already synchronized (see, e.g., Torres, Andersen, \& Gimenez 2010).
\end{itemize}

\subsection{Erroneous $a/R_1$?}

It should be noted that a smaller value for $a/R_1$, closer to \strike{6}\ron{6.2} than its current estimate of $7\pm0.1$, would resolve the discrepancy between ELV and DB amplitudes for $M_1 \simeq 2.5$ $M_\odot$. If this lower value where accurate, then \strike{$q\simeq 0.08$}\ron{$q\simeq 0.11$}, as would be estimated using the DB amplitude alone. However, the temperatures of the primary and secondary would be reduced, according to the ILL model, with lower values of $a/R_1$.  In particular we find, \strike{$T_1 \simeq 7500$ K for $a/R_1 = 6.0$}\ron{$T_1 \simeq 8400$ K for $a/R_1 = 6.15$}.  Additionally, the inverse correlation between $R_2/R_1$ and $a/R_1$ (see Fig.~4) suggests a relatively {\em larger} white dwarf radius for a smaller $a/R_1$.

We have taken care to correctly model the transit and eclipse light curves (see \S 2) but note that, given the extreme correlation between inclination and $a/R_1$ (see Fig.~4), small errors, model inaccuracies, or non-uniform data sampling may lead to systematically inaccurate values for $a/R_1$ that vary greatly in an absolute sense. 

We remark that the $\sim$40 data points occurring during the ingress or egress phases of the secondary eclipse carry the overwhelming majority of the statistical weight in determining $a/R_1$, $i$ and $e \cos(\varpi)$.  While the MCMC algorithm (\S 2.2) will likely yield a robust measure of parameter precision and correlation, the accuracy of the most-likely value is affected by correlated noise and the (sparse) distribution of data points during eclipse ingress or egress.  Compiling future {\em Kepler} data for KHWD3 will most likely resolve any current statistical biases. 

\section{Inferred Mass of the White Dwarf}

Recall from \S3 that, depending on whether our analysis emphasized the ELV and illumination amplitudes or the Doppler boosting and illumination amplitudes, we find that $q = 0.150 \pm 0.006$ or \strike{$q \simeq 0.070 \pm 0.01$}\ron{$q = 0.105 \pm 0.008$}, respectively (see \S3 for details).  For each of these values of $q$, we would infer a white dwarf mass of
\begin{eqnarray}
M_2  & = & (0.37 \pm 0.08) \left(\frac{M_1}{2.5 \,M_\odot}\right) \ron{M_\odot}~~~{\rm ELV ~\&~ILL}\\
M_2 & = &\ron{(0.26 \pm 0.04)} \left(\frac{M_1}{2.5\,M_\odot}\right) \ron{M_\odot} ~~~{\rm DB ~\&~ILL}
\end{eqnarray}
(see Table 2).  We discuss the consequences of these values for the binary stellar evolution of the system in the next section.

\section{System Progenitors}

The presence of a white dwarf in KHWD3, coupled with the short orbital period (3.273 days), indicate that there was necessarily a phase of mass transfer/loss during the prior evolution of this binary.  One obvious constraint on the primordial binary is that it must have had a total mass, $M_p+M_s$, equal to at least the current mass of the binary $M_1+M_2 \simeq 2.6 \,M_\odot$, where $M_p$ and $M_s$ are the initial masses of the primordial primary (the white dwarf progenitor) and the secondary, respectively.  As is readily evident, and we explore below, this implies that the mass of the primordial primary was greater than $1.5\,M_\odot$, and therefore had a radiative envelope at the time when mass transfer to the primordial secondary commenced.  In turn, this indicates that the mass transfer from the primary to the secondary took place, at least initially, on a thermal timescale\ron{.}\strike{, thereby quickly halting the growth of the He core that had already developed.} 

Given the two different possible masses for the white dwarf (see \S4), there are two slightly different evolutionary scenarios for the formation of the white dwarf.  We discuss each of these in turn.

\subsection{Primordial Primaries with $M \lesssim 2.2 \, M_\odot$}

Even if the donor star is more massive than the accretor, the mass transfer can still proceed quite stably if the donor has a radiative envelope.  The mass ratio of the primordial progenitor binary, $q_{\rm prog} \equiv M_p/M_s$, up to which the mass transfer is stable, depends on the masses and the orbital period when mass transfer commences, but can exceed $q_{\rm prog} \sim 2$.  For stars with initial mass $\lesssim 2.2 \,M_\odot$, a degenerate He core develops and there is a nearly unique relation between the radius of the evolving star and the core mass.  This leads to a tight relation between the mass of the remnant white dwarf, once the envelope of the primary has been completely transferred to the secondary and/or lost from the system, and the orbital period:
\begin{equation}
P_{\rm orb} \simeq 4.6 \times 10^6 \, \frac{M_{\rm wd}^{9}}{(1+25M_{\rm wd}^{3.5}+29M_{\rm wd}^6)^{3/2}} ~~ {\rm days}
\end{equation}  
where $M_{\rm wd}$ is in units of $M_\odot$ (see, e.g., Rappaport et al.\,1995; {Ergma 1996; Tauris \& Savonije 1999; Lin et al.\,2010}).  If we solve this non-linear equation for the value of $M_{\rm wd}$ that matches the current orbital period of 3.273 days, we find:
\begin{equation}
M_{\rm wd} \equiv M_2 \simeq 0.21 \pm 0.02 ~M_\odot
\end{equation}  
where the estimate in $M_{\rm wd}$ is uncertain by {$\sim$10\%} (Rappaport et al.\,1995).  This mass range is marginally consistent with the lower of the two possible masses measured for the white dwarf in KHWD3 \ron{($0.26 \pm 0.04 \, M_\odot$)}.

Another major issue that needs to be addressed regarding the white dwarf is how to explain its large radius ($R_2 = 0.15 \pm 0.01 \,R_\odot$) and high effective temperature (\strike{$T_2 = 14,000 \pm 350$ K}\ron{$T_2 = 14,500 \pm 300$ K}).  In particular, the radius is approximately $7$\ron{$-9$} times the radius of a degenerate He star of the same mass.  According to cooling models of He white dwarfs formed in an analogous manner (i.e., after losing its envelope to a companion star; Hansen \& Phinney 1998; Nelson, Dubeau, \& MacCannell 2004; Ph.\,Podsiadlowski, private communication 2010) the radius can remain this bloated for a considerable period of time for very low-mass white dwarfs.  The degree to which this is possible depends on the amount of residual H-rich atmosphere the core retains after losing its envelope.  In particular, a He white dwarf of 0.21 $M_\odot$ with a H-rich atmosphere of 0.01 $M_\odot$ can remain this large and hot for $\sim$150 Myr, while  a 0.19 $M_\odot$ He dwarf with a similar H-rich atmosphere can remain thermally bloated and hot for up to 300 Myr.  Even in the case of 0.21 $M_\odot$ white dwarf, 150 Myr is a non-negligible fraction of the nuclear evolution time of the $\sim$$2.3\,M_\odot$ parent star.  Degenerate He white dwarfs with masses $\gtrsim 0.25 \,M_\odot$ would not remain hot and thermally bloated for nearly long enough to have a plausible probability of catching them in such a state.  \ron{Therefore, the lower limit of the measured range of $0.26 \pm 0.04 \, M_\odot$ would be strongly preferred in this scenario.}
        
\subsection{Primordial Primaries with $M \gtrsim 2.2 \, M_\odot$}

Stars with mass $\gtrsim 2.2 \, M_\odot$ do not evolve with degenerate cores and their radius does not follow the radius--core mass relation discussed above.  In a close binary system, such stars are {less likely} to produce a remnant He core with a mass as low as \strike{$\sim$0.18 $M_\odot$}\ron{$\sim$0.19-0.22 $M_\odot$}.  However, they could produce a white dwarf that matches the more massive of our allowed solutions, namely $\sim$$0.37 \pm 0.08 \, M_\odot$.  Such non-degenerate He stars, once the envelope of the primordial primary has been removed, would be on the He-burning main sequence (as long as $M_{\rm core} \gtrsim 0.31\,M_\odot$), and they could be quite luminous. 

An analytic fitting function for the luminosity of naked He burning stars is:
\begin{equation}
L_{\rm He}= \frac{1.53 \times 10^4 ~M_{\rm He}^{10.25}}{M_{\rm He}^9 + 29.5 M_{\rm He}^{7.5}+31.2M_{\rm He}^6+0.047} ~~L_\odot
\end{equation}
(Hurley, Pols, \& Tout 2000; hereafter ``HPT'').  The observed luminosity of the white dwarf in KHWD3 is $L_2 \simeq 0.80 \pm 0.13 \,L_\odot$.  The He star mass that corresponds to this luminosity, according to the above fitting formula, is down near the end of the He-burning main sequence with $M_{\rm He} \lesssim 0.32 \, M_\odot$.  The radius and effective temperature of such a He burning star would be $\sim$$0.055 \,R_\odot$ and $\sim$26,000 K, respectively (HPT).  The corresponding measured radius and effective temperature of the hot white dwarf in KHWD3 is about 2.5 times larger in radius, and about half as high in temperature.  This could be nicely accommodated (and produce roughly the same luminosity) if the He star has a modest residual hydrogen-rich atmosphere (i.e., $\gtrsim 0.02 \,M_\odot$; see Han et al.\,2002; Han et al.\,2003).  The nuclear lifetime of such a He-burning star, i.e., near the end of the He-burning main sequence, is $\sim$$10^9$ yr.  Such a star would be at the low-mass end of what are known as subdwarf B (sdB) stars (see, e.g., Han et al.\,2003, their figures 15 and 17).  

\subsection{Possibility of a Prior Common-Envelope Phase}
    
We note that unstable mass transfer leading to the current system probably could not have resulted in a common envelope (``CE'') phase leading to the successful ejection of the CE.  The ejection of a common envelope around such a low-mass core would have resulted in a post-CE orbital period that is considerably shorter than 3 days.  The only way in which a 3-day post-CE binary would ensue is if the pre-CE orbital period were extremely long -- in which case the white dwarf mass would be much higher than is observed.  For an impressive study of such post common-envelope systems see the work by Parsons et al. (2010).

\subsection{Constraints on $M_p$, $M_s$ and $P_{\rm orb, init}$}

We explore here what the range of possible and likely parameters the primordial progenitor binary could have been.  We know three independent binary parameters of the current system: $M_1$, $M_2$, and $P_{\rm orb}$.  The appropriate application of conservation of mass and angular momentum yield two constraints -- insufficient to uniquely identify, by themselves, the initial system parameters.  However, there are two other constraints, discussed here, that are sufficient to define a relatively narrow range in parameter space for the primordial binary. 

As mass is transferred from the primary to the secondary, a fraction of it, $\beta$, will be retained by the secondary, and the remainder will be lost from the system.  The ejected mass will carry away an average specific angular momentum which we denote as $\alpha$ which is in units of the specific angular momentum of the binary.  Unfortunately, we do not know {\em a priori} either $\alpha$ or $\beta$, but we can make an educated guess about the former.  In terms of the quantities we have already defined earlier, we can write the mass retention fraction as:
\begin{equation}
\beta = \frac{M_1-M_s}{M_p - M_2}
\end{equation}
Conservation of angular momentum then yields a relation between the orbital periods before ($P_i$) and after ($P_f$) the mass transfer phase:
\begin{equation}
P_i = P_f \left(\frac{M_p+M_s}{M_1+M_2}\right) \left(\frac{M_p}{M_2}\right)^{C_1} \left(\frac{M_s}{M_1}\right)^{C_2}
\end{equation}
where the powers, $C_1$ and $C_2$ are defined as
\begin{eqnarray}
C_1 & = & 3 \alpha (1 - \beta) - 3 \\
C_2 & = & -3 \alpha (1 - \beta)/\beta - 3
\end{eqnarray}
(see, Podsiadlowski et al.\,1992; Rappaport, Podsiadlowski, \& Horev 2009).  In the scenario discussed above for forming a white dwarf of mass \strike{$\sim$0.2 $M_\odot$}\bon{$\sim$0.25} $M_\odot$, we can constrain $P_i$ to be shorter than $P_f = 3.273$.  In the case where a more massive white dwarf formed from a primary star with $M_p \gtrsim 2.2 \,M_\odot$, the initial orbital period could, in principle have been longer than 3.273 days, and the orbit could have subsequently shrunk during mass transfer.  However, for substantially longer initial $P_{\rm orb}$, the core mass would likely exceed $\sim$0.4 $M_\odot$.

Another constraint that we can impose is that the initial orbital separation should be large enough so that not only can both the primordial primary and secondary fit within their respective Roche lobes, but that there is sufficient room for the primary to evolve a substantial He core before the mass transfer commences.  This constraint can be written as:
\begin{equation}
R_p(M_p) = \xi f(q) a = \xi f(q)\left[G(M_p+M_s)\right]^{1/3} \left(\frac{P_i}{2 \pi}\right)^{2/3}
\end{equation}
where $\xi$ is the fraction of its Roche lobe that is filled by the primordial primary of radius $R_p$; $f(q)$ is the ratio of the Roche-lobe radius to orbital separation, $a$, which depends only on the mass ratio, $q$; and $P_i$ is the orbital period before the mass transfer commences.  We take as a somewhat arbitrary but quite reasonable constraint: $\xi \lesssim 1/2$.

\begin{figure}[h!]
\begin{center}
\epsscale{1.18}
\plotone{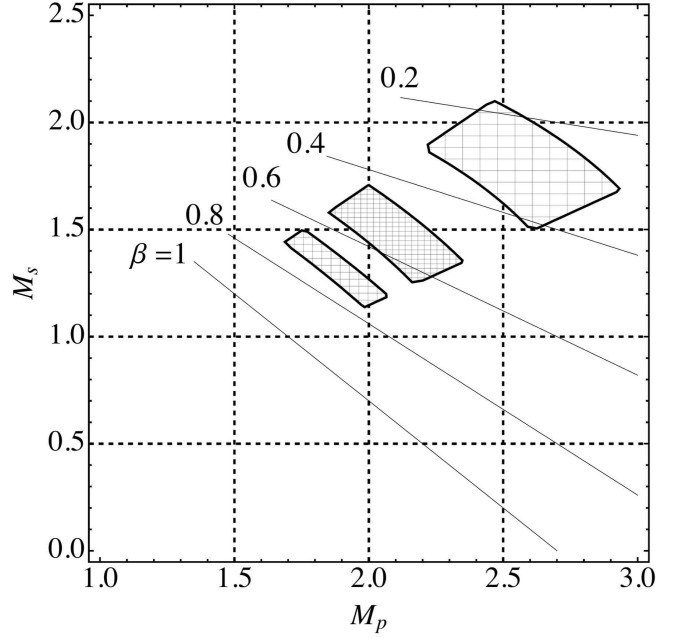}
\caption{Regions in the plane of $M_p$ and $M_s$ where the progenitor binary that produced KHWD3 could have originated. The three shaded regions correspond to the angular momentum loss parameters $\alpha=$ 0.75, 1.0, and 1.50, from top right to lower left.  The upper left and lower right boundaries of each region are determined by $\tau_1/\tau_2 = 0.6$ and 0.2, respectively. The upper right boundaries are set by $P_{\rm orb} = 3.273$ days, while the lower left boundaries are set by allowing the primordial primary to fill no more than half of its Roche lobe. The contours are of constant mass retention fraction, $\beta$.}
\label{fig:regions}
\end{center}
\end{figure}

A final constraint is the ratio of the thermal and nuclear timescales of the primary to those of the secondary.  We have shown (Rappaport et al.\,2009) that these both scale roughly as
\begin{equation}
\frac{\tau_1}{\tau_2} \simeq \left(\frac{M_p}{M_s}\right)^{-2.6}.
\end{equation}
We require that this ratio be smaller than $\sim$0.6 so that, while the original primary evolved, the secondary (the current primary) did not evolve substantially off the main sequence by the current epoch.  On the other hand, we simultaneously require that this ratio be larger than about 0.2 so that the thermal timescales of the two primordial stars not be too dissimilar.  If they were very different, then it is likely that the mass transfer would have become dynamically unstable.  In summary, we require that $0.2 \lesssim \tau_1/\tau_2 \lesssim 0.6$.

We adopt a nominal value for the specific angular momentum loss parameter, $\alpha =1.0$; however, we also consider values of $\alpha = 0.75$ and 1.5.  This is a reasonable set of bounds since, for example, the specific angular momentum parameter, $\alpha$, at the $L2$ point where matter is likely to escape from the binary system, ranges from 1.08 to 1.56 for a mass ratio covering the rather wide range of $0.1 \lesssim M_s/M_p \lesssim 2$.

When we combine the constraints imposed by conservation of mass, angular momentum, ratio of timescales, and initial Roche-lobe filling factors we find an acceptable region in initial parameter space that is shown in Fig.\,\ref{fig:regions}.

\begin{figure}[t]
\begin{center}
\epsscale{1.2}
\plotone{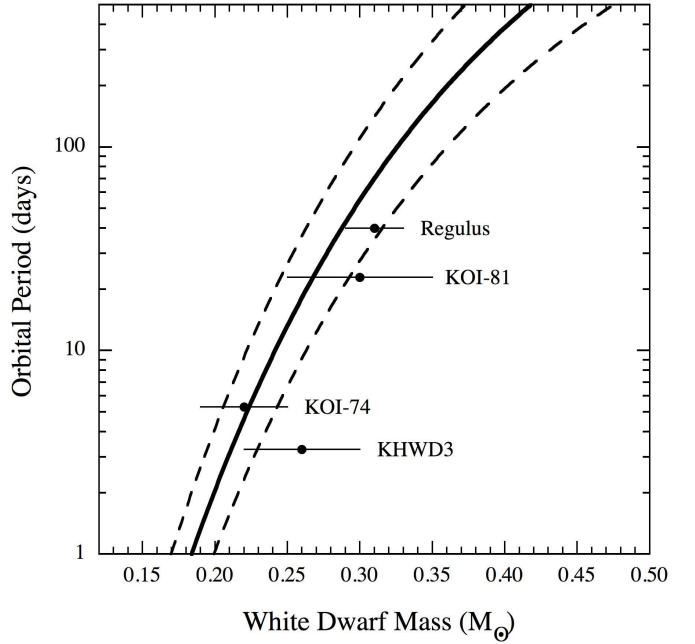}
\caption{Comparison of the four known systems with white dwarfs that are relics of stable Roche-lobe overflow.  The plotted show the orbital period vs. the white dwarf mass for KHWD3, KOI-74, KOI-81, and Regulus.  The heavy curve is the theoretically derived $P_{\rm orb}(M_{\rm wd})$ relation, going as {$\sim$$M_{\rm wd}^9$ for the lowest mass white dwarfs}\bon{, while the pair of dashed curves marks the estimated theoretical uncertainties (see text for details and Rappaport et al.\,1995).}}
\label{fig:HWD}
\end{center}
\end{figure}

\subsection{Comparison of KHWD3 with KOI-74, KOI-81, and Regulus}

We end our discussion of the prior evolutionary history of KHWD3 by comparing this system with three other white dwarfs in close binaries that also likely formed during a phase of stable Roche-lobe mass transfer.  These are KOI-74, KOI-81, and Regulus with white dwarfs in orbital periods of 5.3 days, 23 days, and 40 days, respectively (Rowe et al.\,1010; van Kerkwijk et al.\,2010; Gies et al.\,2008; Rappaport et al.\,2009).  The measured white dwarf masses in these systems are $0.22 \pm 0.03$, $\sim$0.3, and $0.30\pm 0.02$ $M_\odot$. Including the lower-mass solution for KHWD3 (i.e., \strike{$0.18 \pm 0.03 \, M_\odot$}\ron{$0.26 \pm 0.04 \, M_\odot$}), we find a \strike{secular progression of masses with increasing $P_{\rm orb}$}\ron{modestly significant correlation between masses and} increasing $P_{\rm orb}$.  This correlation is plotted in Fig.\,\ref{fig:HWD} where the solid curve is the theoretically expected relation for progenitor stars of mass $\lesssim 2.2\,M_\odot$ (see eq.\,29 above).  The white dwarf masses are not very well determined, and there is only a small spread in all their masses by a factor of \strike{$\sim$2}\ron{$\sim$1.5}.  Nonetheless, given the extremely steep theoretical dependence of $P_{\rm orb}$ on the mass of the white dwarf, the known systems \strike{seem to fit this relation reasonably well}\ron{are roughly consistent with this relation}.  More such systems, with even better determined white dwarf masses, will be needed \ron{ in order} to supplement the kind of information currently being provided by radio pulsar systems (Rappaport et al.\,1995; Tauris 1998; Thorsett \& Chakrabarty 1999).

\section{Summary and Conclusions}

In this work we have utilized public {\em Kepler} data to identify and study in detail a third hot white dwarf in a close binary orbit.  The system is formally known as KIC 10657664, and we have given it the shorthand name ``KHWD3''.  The system is comprised of a white dwarf with \strike{$T_{\rm eff} \simeq 14,000$}\ron{$T_{\rm eff} \simeq 14,500$} K orbiting an A star of mass $\sim$2.3 $M_\odot$.  The white dwarf is extremely thermally bloated with a radius of $0.15\,R_\odot$, some $7-10$ times larger than its degenerate radius.  

We have combined the transit and eclipse data as well as the low-amplitude, out-of-transit/eclipse light curve to infer all the system parameters.  The large size of the white dwarf, coupled with its high effective temperature, produces relatively large eclipse and transit depths ($\sim$2\% and 0.7\%, respectively).  The out-of-transit/eclipse periodic light curves are of much smaller amplitude (in the range of $\sim$$1-5$ parts per $10^4$) and are interpreted as being due to Doppler boosting, mutual illumination, and ellipsoidal light variations (due to tidal distortions).  The deduced system parameters are summarized in Table 2.  

The only ambiguity in the determination of the system parameters lies in the mass of the white dwarf.  Our analysis leads to two \ron{somewhat} distinct, and currently unresolvable, possible solutions: \strike{$0.18 \pm 0.03 \, M_\odot$}\ron{$0.26 \pm 0.04 \, M_\odot$} or $0.37 \pm 0.08 \, M_\odot$.  The difficulty in distinguishing between these two possibilities for the white dwarf mass lies in the fact that both the Doppler effect and the ELV both basically determine the mass ratio, and these two amplitudes in the light curve produce somewhat inconsistent results \ron{in this ratio} (at the \strike{2.2}\ron{$\sim$2.8} $\sigma$ level).  However, it is important to note that the remainder of the important system parameters are well determined, in spite of this particular ambiguity.

We have also briefly explored what the possible progenitors of this system might have been.  The system almost certainly formed after a phase of mass transfer from the primordial primary star (now the white dwarf) to the secondary star (now the A star primary of the system).  If the primordial primary had a mass less than $\sim$2.2 $M_\odot$, its core mass would directly determine the final orbital period (3.273 days), which predicts the current white dwarf mass to be { $0.21 \pm 0.02 \,M_\odot$,} in \strike{excellent}\ron{modest} agreement with the lower of our two solutions.  Such a low-mass white dwarf \ron{(i.e., $0.19-0.22 \,M_\odot$)} could remain hot and bloated for a substantial fraction (i.e., $\gtrsim$ 10\%) of the lifetime of the current primary A star.  If the primordial primary had a mass $\gtrsim 2.2\,M_\odot$, then it is likely that our solution for a more massive white dwarf is the correct one.  In this case, however, the white dwarf would almost certainly be undergoing nuclear burning at the current epoch.  This, in turn, would imply that the mass would have to be low ($\simeq 0.31-0.32\,M_\odot$), i.e., near the end of the He-burning main sequence, in order to explain a luminosity of $\lesssim 1 \,L_\odot$.  

Likely primordial stars of mass $M_p \simeq 2.2 \,M_\odot$ and $M_s \simeq 1.4 \,M_\odot$ (see Fig.\,\ref{fig:regions}) would imply a highly non-conservative phase of mass transfer with perhaps half the transferred mass being lost from the system.  For this particular set of illustrative primordial binary masses, the mass ratio is sufficiently large to possibly explain the high mass loss rate.  We have compared KHWD3 with the two other hot white dwarfs discovered with {\em Kepler} as well as the Regulus system.  The initial system masses for Regulus (see Rappaport et al.\,2009) were $M_p \simeq 2.1 \,M_\odot$ and $M_s \simeq 1.7 \,M_\odot$, which are seemingly not too different than for KHWD3.  However, the final orbital period of the white dwarf in the Regulus system is 40 days, which is considerably longer than the 3.3-day period of KHWD3.  The difference likely lies in the fact that in the Regulus system the mass transfer was considerably more conservative (leading to a much wider orbit; see Eqn.\,33).  This is quite plausible given that the initial mass ratio of $\sim$1.2 for Regulus was significantly closer to unity than for KHWD3 which may have been closer to $\sim$1.6.  

\acknowledgments
We thank Dave Latham, Al Levine, and Leslie Rogers for helpful discussions.  \bon{We appreciate a careful reading of the manuscript by Tsevi Mazeh and an anonymous referee.  We are grateful to Marten van Kerkwijk for pointing out the importance of using proper stellar atmosphere models to compute the Doppler boosting prefactors.  We thank Frederic Pont for suggesting the correlated noise analysis.}  J.C. acknowledges NASA Origins of Solar Systems grant no.\,NNX09AD36G for partial support of this work.  D.F. is grateful for funding through a Michelson Fellowship, which is supported by NASA and administered by the NASA Exoplanet Science Institute.  J.C. and D.F.  acknowledge partial support for this work was provided by NASA through Hubble Fellowship grant \#HF-51267.01-A and \#HF-51272.01-A awarded by the Space Telescope Science Institute, which is operated by the Association of Universities for Research in Astronomy, Inc., for NASA, under contract NAS 5-26555.

\end{document}